\documentclass[aps,prl,twocolumn,superscriptaddress,longbibliography,nobibnotes]{revtex4-2}
\usepackage[colorlinks=true,linkcolor=blue,citecolor=blue,urlcolor=blue]{hyperref}
\usepackage{inputenc,amsfonts,amsmath,amsthm,amssymb,amsmath,amscd,graphicx,mathrsfs,braket,xcolor,dsfont,physics,comment}

\def\mO{\mathcal{O}}

\renewcommand\vec{\boldsymbol}

\usepackage{ulem}

\begin{document}
\raggedbottom
\title{Designing (higher) Hall crystals}
\author{Nisarga Paul}
\email{npaul@mit.edu}
\affiliation{Department of Physics, Massachusetts Institute of Technology, Cambridge, MA,
USA}
\author{Gal Shavit}
\affiliation{Department of Physics and Institute for Quantum Information and Matter,
California Institute of Technology, Pasadena, CA, USA}
\affiliation{Walter Burke Institute of Theoretical Physics, California Institute of Technology, Pasadena, CA, USA}
\author{Liang Fu} 
\affiliation{Department of Physics, Massachusetts Institute of Technology, Cambridge, MA,
USA}

\begin{abstract}
We introduce a novel platform for realizing interaction-induced Hall crystals with diverse Chern numbers $C$. This platform consists of a two-dimensional semiconductor or graphene subjected to an out-of-plane magnetic field and a one-dimensional modulation, which can be realized by moir\'e or dielectric engineering. We show that interactions drive the system to spontaneously break the residual translational symmetry, resulting in Hall crystals with various $C$ (including $|C|>1$). Remarkably, these phases persist across \textit{continuous} ranges of filling and magnetic field, and the global phase diagram can be understood in a unified manner. 
\end{abstract}

\maketitle
\textit{Introduction.--- } Two-dimensional electronic systems in a strong magnetic field display a rich interplay between charge density wave (CDW) and quantum Hall phases. In the lowest Landau level, Wigner crystals compete with fractional quantum Hall (FQH) states at low fillings\cite{Stormer1999Mar,Tsui2024Apr,Fogler1996Jul}. Even more interesting is the theoretical  prediction that crystalline order and the quantized Hall effect may coexist, leading to a novel quantum phase of matter referred to as the Hall crystal \cite{Tesanovic1989Apr}. 
\par 

Besides exhibiting a topologically quantized Hall effect, the most striking fact about Hall crystals, first noted by Halperin et al. \cite{Halperin1986Aug}, is that the number of electrons per unit cell can vary continuously, as can the number of flux quanta per unit cell. However, they satisfy a definite relation, the Diophantine relation \cite{Halperin1986Aug}, which relates them to rational topological invariants such as the many-body Chern number $C$. The Hall conductivity of the Hall crystal (pinned by an infinitesimal potential) is  $\sigma_H = \frac{e^2}{h} \pdv{\bar \rho}{\rho_{\Phi}} = \frac{e^2}{h}C$ \cite{Thouless1982Aug,Niu1985Mar}. 
\par 
The study of Hall crystals was initially motivated by the Wigner-crystal-to-FQH transition at low fillings of the lowest Landau level (LL) \cite{Kivelson1986Feb,Halperin1986Aug,Tesanovic1989Apr}. Later, Hartree-Fock studies revealed crystalline electron 
phases in partially filled higher LLs  \cite{Koulakov1996Jan,Moessner1996Aug,Rezayi1999Aug,Shayegan2000Feb}, which also exhibit a quantized Hall effect due to the completely filled LLs. Recent observations of quantum anomalous Hall (QAH) states in moir\'e materials~\cite{Cai2023Oct,Park2023Oct, Lu2024Feb} have sparked great interest in ``anomalous'' Hall crystals~\cite{Song2024Mar, Sheng2024Feb, Dong2023Nov, Dong2023Nov2}, requiring no external magnetic field. 
Indeed, evidence of a QAH crystal with $C=1$ has been observed at fractional filling of twisted bilayer-trilayer graphene \cite{Su2024Jun}.       
\par 
\begin{figure}
    \centering
\includegraphics[width=\linewidth]{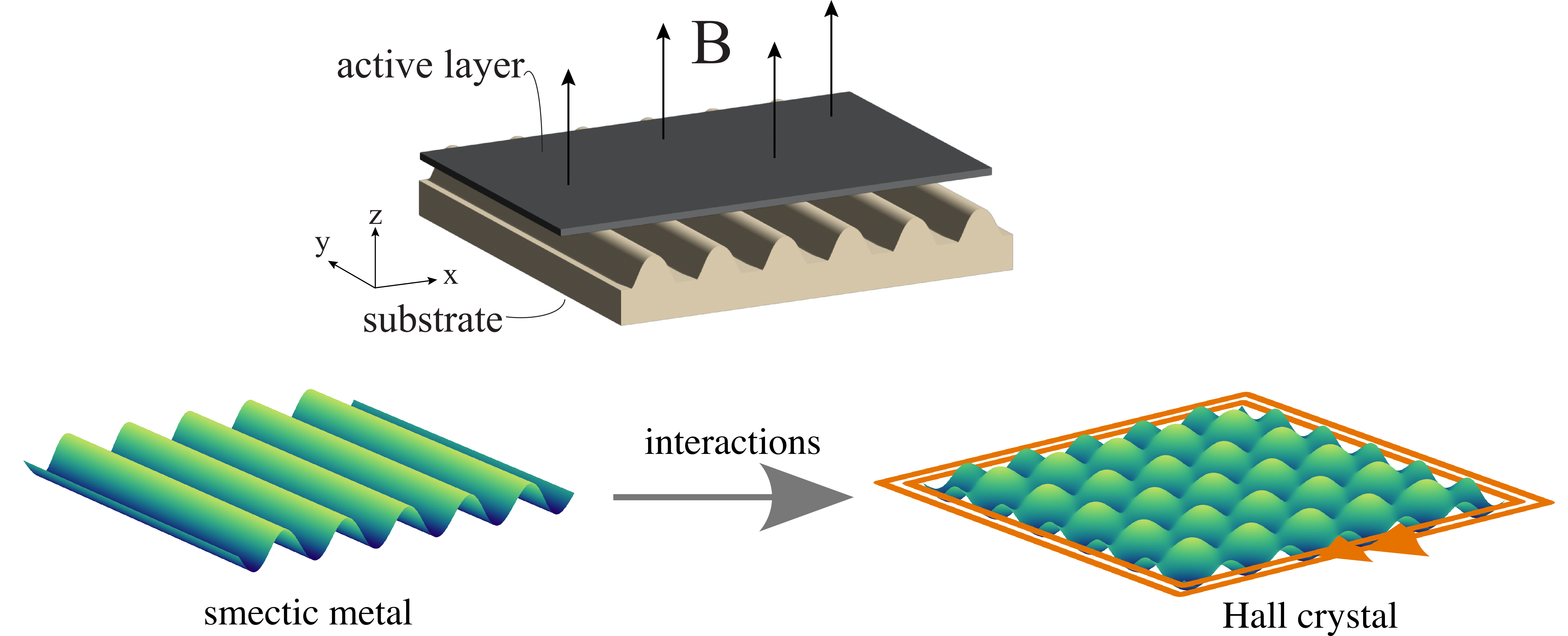}
    \caption{(Top) An active semiconductor or graphene layer with free charge carriers subject to a uniform field $B\hat z$ and a 1D modulation $V(x)$ from a patterned substrate. (Bottom) Electron interactions drive CDW formation along the $y$ direction resulting in Hall crystals with various Chern numbers.}
    \label{fig:device}
\end{figure}
Existing studies of (anomalous) Hall crystals have largely focused on states with unit many-body Chern number $|C| = 1$ arising from a partially filled Landau level or Chern band.  An intriguing research direction is the study of Hall crystals with $|C| > 1$, which we dub ``higher'' Hall crystals. Such states have not been observed in the ordinary quantum Hall effect and may exhibit interesting new physics when interactions are present \cite{Moller2015Sep,Liu2021Jan,Lee2018Jul}. The remarkable tunability and customizability of two-dimensional materials further motivate the search for such phases.
\par

In this work, we introduce a realistic and tunable platform for Hall crystals with various many-body Chern numbers. By analytically controlled theoretical analysis, we show that (higher) Hall crystals can reliably emerge from electron interactions in an engineered setting, with tunable Hall conductivity and lattice constant.  
\par 
Our platform consists of a 2D semiconductor (such as graphene) subject to a \textit{one}-dimensional superlattice potential $V(x)$ and an out-of-plane $B$ field, as shown in Fig. \ref{fig:device}. We show that the Coulomb interaction in a periodically modulated LL induces various CDW instabilities that spontaneously break the continuous translation symmetry along $y$ direction, and the resulting {\it two}-dimensional electron crystal states have diverse Chern numbers leading to an intricate phase diagram as a function of electron density and magnetic field. 
\par 

\textit{Setup and model.--- } We start by considering Landau levels (LLs) of a two dimensional electron system that are subject to a one-dimensional periodic potential $V(x)$ with period $a_0$. We decompose the potential into harmonics as $V(x) = \sum_i V_i \cos(q_i x)$ and consider strong magnetic fields $|V_i| \ll \omega_c= eB/m$. To first order in $V_i/\omega_c$, LL mixing can be neglected and the energy spectrum of the $n$-th Landau level becomes dispersive as a function of $k_y$: 
\begin{eqnarray}
E_{n}(k_y)=  \sum_i V_i e^{-q_i^2\ell^2/4}L_n(q_i^2\ell^2/2) \cos(q_i k_y \ell^2),     
\end{eqnarray}
with $\ell = 1/\sqrt{B}$ the magnetic length ($\hbar = e =1$), while remaining flat as a function of $k_x$. This quasi-one-dimensional energy dispersion is a consequence of the locking between electron position $x$ and momentum $k_y$ in a Landau level.  
Indeed, in the high field limit $\ell \rightarrow 0$, $E_{n}(k_y)= V(x= k_y \ell^2)$ exactly traces the potential landscape. 
\par
In the presence of electron interactions, the LL-projected Hamiltonian can naturally written in the Landau gauge basis, taking the form (dropping the $y$ subscripts from now)
\begin{subequations}
    \begin{align}
        H &= H_0 + H_1 \\
        H_0 &= \sum_{k} E_n(k) c_{n,k}^\dagger c_{n,k} \\
        H_1 &= \sum_{k_1,k_2,k_3,k_4} U_{k_{1},\ldots, k_{4}} c_{n,k_{1}}^\dagger c_{n,k_{2}}^\dagger c_{n,k_{3}} c_{n,k_{4}} 
    \end{align}
\end{subequations}
where $c_{n,k}^\dagger$ is the creation operator for the LL orbital in the Landau gauge $\varphi_{n,k}(\vec r) = \frac{1}{\sqrt{L_y\ell\sqrt{\pi}2^nn!}} e^{-ik y}H_n((x-k\ell^2)/\ell)e^{-(x-k\ell^2)^2/2\ell^2}$ and $U_{k_{1},\ldots, k_{4}}$ is an interaction matrix element. 
\par
\begin{figure}
    \centering \includegraphics[width=\linewidth]{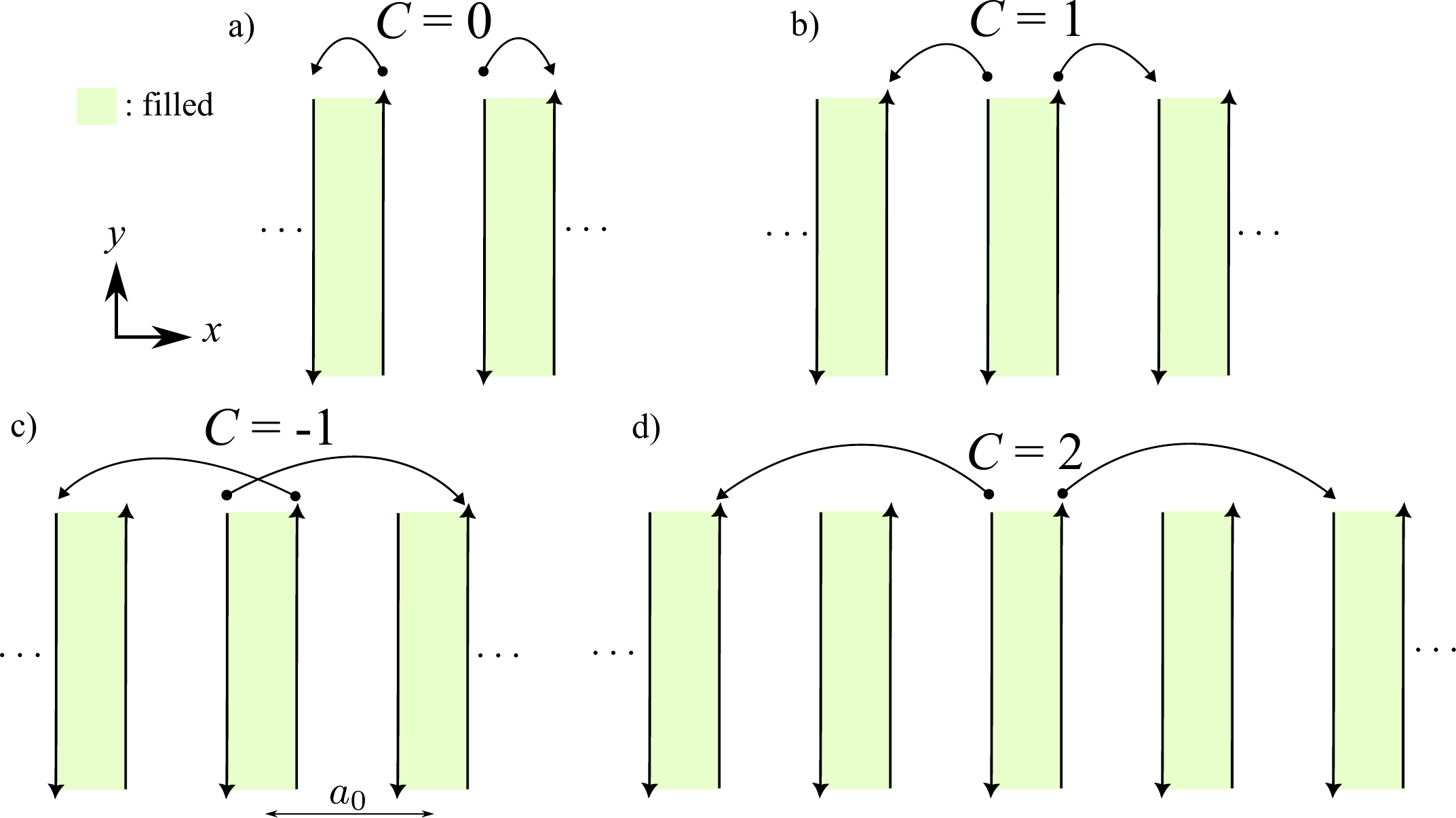}
    \caption{Coupled chiral wires with four-fermion interactions illustrated as correlated hoppings between wires, whose strong coupling phases have the indicated Chern numbers. }
    \label{fig:wires}
\end{figure}

 
We now analyze our setup microscopically. For the purpose of a rigorous analytical treatment, we assume that the interaction energy scale is small compared to the bandwidth of the modulated LL: $W\sim |V|$ (which is in turn much smaller than the cyclotron energy $\omega_c$).  In this weak interaction regime, let us first consider the effect of energy dispersion $H_0$.   
When the $n$-th LL is partially filled, low-energy electron states are one-dimensional chiral modes associated with equipotential lines of $E_n(k)=\mu$ with $\mu$ the chemical potential, forming an array of 1d chiral fermion wires along the $y$ direction (see Fig. \ref{fig:wires}). These wires have alternating chiralities and spacings $\delta a_0$ and $(1-\delta) a_0$, and $\delta$ increases with the LL filling factor $\nu$ from $0$ to $1$. For example, in the case of a sinusoidal potential $V(x)=V\cos(2\pi x/a_0)$, $\delta=\nu$, the positions are $x_I =a_0(\lfloor(I-1)/2\rfloor+1/2+(-1)^I \nu/2)$ \footnote{This also holds for any periodic $V(x)$ with a single local maximum and local minimum per unit cell coincident with those of $\cos(2\pi x/a_0)$} and the velocities are $v_I = \pdv{E_n}{k_{y}}|_{k_{I}} \equiv (-1)^I v$ (with $v>0$). Even (odd) wire index corresponds to a positive (negative) velocity along the $y$ direction. 
\par 
The low-energy theory then consists of chiral fermions $\psi_I = \int_{k_{I}-\Lambda}^{k_{I}+\Lambda} \frac{dk}{2\pi} e^{-i ky}c_{n,k}$, where $I=1,\ldots 2N$ and $N\to \infty$ in the thermodynamic limit. 
By retaining only terms involving low-energy chiral fermions $\psi_I$ in $H$, we obtain the effective Hamiltonian 
\begin{equation}
    H_{\text{eff}} = \int dy\, iv_I \psi_I^\dagger \partial_y \psi_I + U_{II'J'J} \psi_I^\dagger \psi_{I'}^\dagger \psi_{J'}\psi_J
\end{equation}
where indices are implicitly summed ($k_{I} \sim I$). Since our setup has translation symmetry along $y$ direction, $H_{\text{eff}}$ satisfies $k$ conservation  
\begin{equation}
    k_{I} + k_{I'} = k_{J} + k_{J'}.
\end{equation}
This, along with charge conservation, 
places important kinematic constraints on the coupled-wire effective theory: allowed interactions in $H_{\text{eff}}$ are correlated hoppings of a pair of opposite-chirality fermions $(I_0, J_0) \rightarrow (I_0-l, J_0 +l) $. They can all be written as 
\begin{equation}\label{eq:I0}
\mO_{I_0,J_0,l} = \int dy \, \psi_{J_0+l}^\dagger \psi_{I_0 - l}^\dagger \psi_{I_0}\psi_{J_0} +\text{H.c.}
\end{equation}
where $I_0<J_0$, $J_0 - I_0 \mod 2 = 1$ and $l\geq 0$. Such terms preserve the center-of-mass position along $x$ direction as required by $y$ momentum conservation.  

The $l=0$ terms, which we refer to as forward scattering, conserve the number of electrons in each wire and have coefficient
\begin{equation}\label{eq:UIJ}
    \mathsf{U}_{IJ} = U_{IJJI}-U_{IJIJ}.
\end{equation}
Importantly, both direct and exchange interactions involving momentum transfer $q\sim 0$ and $k_{J}-k_{I}$ contribute to the forward scattering process $\mathsf{U}_{IJ}$. The $l\neq 0$ terms will be referred to as correlated hoppings.
\par 

{\it CDW orders and Hall crystals---} An array of coupled quantum wires is generically unstable to interactions \cite{Emery2000Sep,Vishwanath2001Jan,Mukhopadhyay2001Jul,Sondhi2001Jan}. Before analyzing the exact order realized for a given set of microscopic parameters, we discuss general features of the ordered phases associated with interactions such as $\mO_{I_0,J_0,l}$. In the thermodynamic limit, $\mO_{I_0,J_0,l}$ and its unit-cell-translated copies induce long-range order (LRO) for the CDW order parameter $\psi_{I_*}^\dagger \psi_{J_*}$, where $I_*$ and $J_*$ are an opposite-chirality pair in $\{I_0-l,I_0,J_0,J_0+l\}$ such that $J_* - I_* \mod 2 = 1$. The resulting CDW has wavevector $Q = |k_{I_*} - k_{J_*}|$ along the $y$ direction. In particular, 
\begin{equation}\label{eq:Q}
    Q = \begin{cases}
        |k_{J_0+l}-k_{I_0}| & l \text{ even}\\
        |k_{I_0}-k_{I_0-l}| & l \text{ odd}
    \end{cases}.
\end{equation}
In general, the resulting CDW is topological. One indication is the presence of $N_{\text{edge}}$ gapless edge modes on a system with a boundary along $x$. Equivalently, $N_{\text{edge}}$ is the number of wires left ``untouched" by by the operator Eq. \eqref{eq:I0} and its translated copies. Some careful counting shows that $N_{\text{edge}}$ takes the values shown in Table \ref{table}. 
\begin{table}
\centering
\begin{tabular}{|c|c|c|c|c|}
\hline
$I_0$ & $l$ &  $N_{\text{edge}} = C$ & $Q\ell^2/a_0 = \Phi_0^{-1}$ \\ \hline\hline
even & even &  $(J_0 - I_0 +l +1)/2$  &  $(J_0 - I_0 +\ell +1)/2 - \nu$ \\ \hline
even & odd&  $-(l-1)/2$ & $(l-1)/2+\nu$ \\ \hline
odd &  even &$-(J_0-I_0+l-1)/2$&  $(J_0-I_0+l-1)/2 +\nu$ \\ \hline
odd & odd &  $(l+1)/2$ & $(l+1)/2-\nu$\\ \hline
\end{tabular}
\caption{Properties of CDW operator $\mO_{I_0,J_0,l}$ (Eq. \eqref{eq:I0}) where $C$ is the Chern number (equal to number of gapless edge modes $N_{\text{edge}}$), $Q$ is the wavevector along $y$, $\ell=$ magnetic length, $a_0=$ period along $x$, and $\nu =$ filling, and $\Phi_0= \#$ flux quanta per unit cell.}
\label{table}
\end{table}

This implies that the CDW can carry a nontrivial Chern number. Indeed, the state is a Hall crystal, as we now show. We employ the Diophantine formula Eq. \eqref{eq:diophantine} relating density, field, and unit cell area in crystalline Hall systems, with $A_0 = (2\pi/Q)a_0$. All these quantities can vary continuously but satisfy the Diophantine relation
\begin{equation}\label{eq:diophantine}
    \bar\rho = C \rho_{\Phi} +\eta A_0^{-1}, 
\end{equation}
where $\bar \rho$ is the electron density, $\rho_{\Phi} = 1/2\pi\ell^2$ is the LL degeneracy per unit area, and in the presence of an infinitesimal potential, the insulating crystal has Hall conductance $\sigma_{xy} = C \frac{e^2}{h}$ by the Streda formula. This implies, for our case, 
\begin{equation}\label{eq:diophantine2}
    \frac{\nu}{2\pi \ell^2} = \frac{C}{2\pi \ell^2} + \eta \frac{Q}{2\pi a_0}.
\end{equation}
$Q$ can be deduced across various cases using Eq. \eqref{eq:Q}, and we record the results in Table \ref{table}. Putting this together, we may observe observe that Eq. \eqref{eq:diophantine2} is satisfied precisely with $\eta = (-1)^{I_0-l-1}$ and 
\begin{equation}
    C = N_{\text{edge}},
\end{equation}
a nontrivial agreement necessitated by the bulk-boundary correspondence. Moreover, we can compactly express the wavevector as  
\begin{equation}\label{eq:Q}
    Q = \frac{a_0}{\ell^2}\eta (\nu-C).
\end{equation}
We further detail the nature of these phases in \cite{supp}. The simplest cases are the $C=0$ and $C=1$ insulators (shown in Fig. \ref{fig:wires}a) which correspond to to $(I_0,J_0,l) = (\text{even/odd}, I_0+1,1)$, respectively. In other words, these correspond to strong backscattering between adjacent wires. Na\"ively, we may expect that a $C=0$ or $C=1$ state is \textit{always} preferred. While this is indeed the case in the limit of tightly localized wires $\ell/a_0 \to 0$ we will show that, surprisingly, various other Hall crystals are stabilized under general assumptions.

\par 

Let us comment on the nature of the CDW phases at special rational conditions. From Eq. \eqref{eq:Q}, it follows that a scenario with both a rational flux per unit cell $\Phi_0 =  a_0/Q\ell^2 $ and rational filling $\nu$ can be achieved. In particular, this could arise for $\Phi_0 = s/t$ and $\nu = r/s$ for integers $r,s,t$. These integers are subject to the constraint $t/s = \eta (r/s-C)$, which is equivalent to the noninteracting Diophantine equation
\begin{equation}
    r = C s + \eta t
\end{equation}
which was shown in Ref. \cite{Thouless1982Aug} to hold for a Landau level subject to a periodic potential $U_1 \cos(2\pi x/a_0) + U_2 \cos(Qy)$ with $U_2\ll U_1$, and precisely $\Phi_0 = s/t$, and $\nu = r/s$. Indeed, $\sigma_H = Ce^2/h$ in this context as well, which famously follows from the Kubo formula \cite{Thouless1982Aug}.

\par 
{\it Methods and results.--- } In order to determine the CDW order for a given set of microscopic parameters, 
we proceed to solve the low-energy model of coupled wires, which describes our modulated LL system at weak interaction (compared to the bandwidth). We will bosonize each wire and apply a perturbative renormalization group analysis. 

\par 

We now treat the system by bosonization: a chiral boson field $\phi_I$ is introduced for each wire via $\psi_I = (2\pi \alpha)^{-1/2} e^{ik_{y,I}y} e^{(-)^I i\phi_I} \gamma_I$, where $\alpha$ is a short-distance cutoff and $\gamma_I$ is a Klein factor \cite{Emery2000Sep,Vishwanath2001Jan,Mukhopadhyay2001Jul,Sondhi2001Jan}. The quadratic action is
\begin{equation}\label{eq:SLL}
    S_0 = \frac{1}{4\pi} \int dtdy\, [\mathsf{K}_{IJ} \partial_t \phi_I \partial_y \phi_J - \mathsf{V}_{IJ} \partial_y \phi_I \partial_y \phi_J],
\end{equation}
where $\mathsf{K}_{IJ} = (-)^{I+1}\delta_{IJ}$ and $\mathsf{V}_{IJ} = v \delta_{IJ} + (1/\pi)\mathsf{U}_{IJ}$. (More details on the bosonization procedure can be found in \cite{supp}.) The action Eq. \eqref{eq:SLL}, describing a sliding Luttinger liquid with forward scattering interactions between wires \cite{Mukhopadhyay2001Jul}, is an RG fixed point parametrized by $\mathsf{V}$. 
\begin{figure}
\centering
\includegraphics[width=0.9\linewidth]{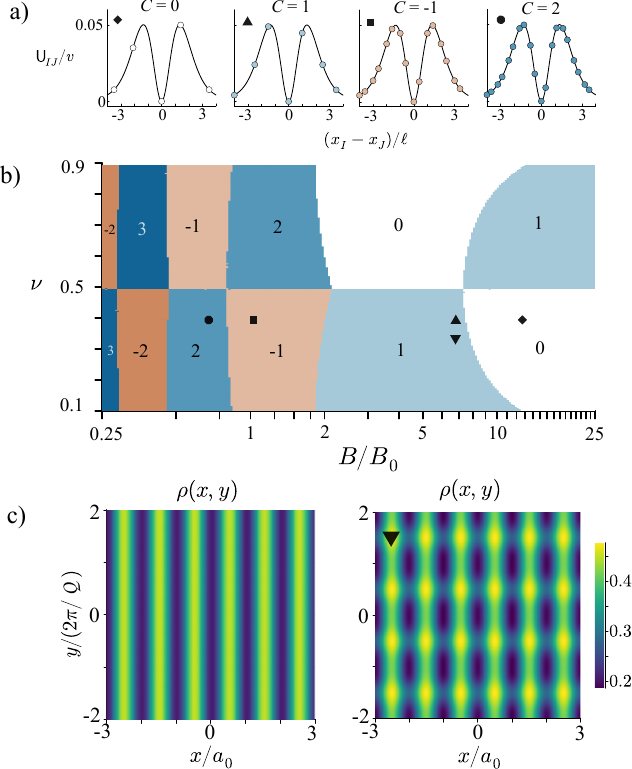}
\caption{(a) Density-density couplings $\mathsf{U}_{IJ}/v$ (dots) and projected lowest LL interactions (line) as functions of interwire separation. The origin, indicated by $0$, is a wire $I_0$ with $I_0$ odd. Symbols indicate $(a_0,\nu)$ values in Fig. \ref{fig:crystals}b (b) Phase diagram of lowest LL modulated by a periodic potential $V(x)$ in a field $B$ with short range Coulomb interactions ($B_0 = \hbar / ea_0^2, \lambda = \ell$). Phases labelled by Chern numbers $C$; each phase has wavevector $Q$ (Eq. \eqref{eq:Q}) and $\eta = -\text{sgn}(C)$. (c) Charge density of a Landau level subject to a periodic potential $U_1\cos(2\pi x/a_0) + U_2\cos(Qy)$, for $U_2 = 0$ (left) and $U_2/U_1 = 0.2$ (right), normalized so that $\bar\rho = \nu$. Filling and flux are $\nu=1/3$ and $\Phi_0=3/2$.}
\label{fig:crystals}
\end{figure}

We proceed to study the effect of the remaining interactions (i.e. correlated hoppings) on these fixed points. A general type of symmetry-allowed perturbation takes the bosonized form 
\begin{equation}\label{eq:Sm}  \int dt  \,\sum_j\mO_{I_0+2j,J_0+2j,l},\quad l\neq 0\end{equation}
where $\mO_{I_0,J_0,l}$ is defined in Eq. \eqref{eq:I0} and has bosonized form 
\begin{equation}\label{eq:mob}
    \mO_{I_0,J_0,l} \simeq \cos(m_I\phi_I)
\end{equation}
with  
\begin{equation}
    (\mathsf{K}\vec m)_I = -\delta_{I,I_0} - \delta_{I,J_0} + \delta_{I,J_0+l}+\delta_{I,I_0-l}.
\end{equation}
It is important to note that Eq. \eqref{eq:Sm} is a sum of local operators that are mutually commuting. \par 
\par 
At the fixed point, if only a single type of correlated hopping Eq. \eqref{eq:Sm} is present, the system develops the corresponding CDW order with wavevector given by Eq. \eqref{eq:Q}. However, because all types of correlated hoppings with different $I_0,J_0,l$ are present for generic interactions, the situation is more complicated. We proceed by searching for the most relevant instability among all types of correlated hoppings, which indicates the strongest ordering tendency at any given set of microscopic parameters.
\par 
To proceed, we take the bare electron-electron interactions to be of Coulomb form, which upon projection to the LL yield matrix elements $U_{II'J'J} = L_y\langle II'|\hat V_C|J'J\rangle$, where $|I\rangle$ denotes a Landau gauge orbital. For $\hat V_C$, we consider both long-range and gate-screened Coulomb interactions with first-quantized form $V_C(\vec r) = (e^2/\epsilon r)e^{-r/\lambda}$. $\epsilon$ is the dielectric constant and $\lambda$ is a screening length. Writing $k_{I'} -k_{I}=q,k_{J'} -k_{J} = p$, we have \cite{supp}
\begin{equation}
U_{II'JJ'} = \frac{\sqrt{2\pi}e^2}{\epsilon }  e^{-\frac{\ell^2(p+q)^2}{8}}\int ds\, K_0\left(\mu(s)\right) e^{-s^2/2}
\end{equation}
for the lowest LL, where $K_0$ is a modified Bessel function of the 2nd kind and
\begin{equation}
    \mu(s) = \frac{\ell}{2}\sqrt{(p+q)^2 + 4\lambda^{-2}}|s+\ell(p-q)/2|.
\end{equation}
The couplings $\mathsf{U}$ depend on $a_0$ and $\nu$ through the positions of the wires.
We show examples of the resulting $\mathsf{U}$ in Fig. \ref{fig:crystals}a.
\par 
Next, we generate a phase diagram by finding the most relevant instability Eq. \eqref{eq:mob} at the fixed point defined by $S_{\text{eff}}[\mathsf{V}]$, with $\mathsf{V}$ set by microscopic parameters, and present the results in Fig. \ref{fig:crystals}. To this end, we need to find scaling dimensions of the operators $\mO_{I_0,J_0,l}$. In the absence of interactions, the scaling dimension is simply $\frac12 \vec m^T\vec m = 2$, as appropriate for a four-fermion operator. When $\mathsf{U}$ is nonvanishing, we may perform a field redefinition $\phi_I' = \mathsf{A}_{I'I} \phi_I$ with $\mathsf{A}$ chosen such to diagonalize the interactions while maintaining the commutation relations; in particular, we may take any $\mathsf{A}$ satisfying $\mathsf{A} \mathsf{V} \mathsf{A}^T = \text{diag}(u_i)$, $\mathsf{A} \mathsf{K} \mathsf{A}^T = \mathsf{K}$, and $\det \mathsf{A} = 1$ \cite{Murthy2020Mar}. It then follows that operator $\mO_{I_0,J_0,l}$ has scaling dimension $\Delta = \frac12 \vec m^T \mathsf{A}^T \mathsf{A} \vec m$.
We note that all the $\mO_{I_0+2j,J_0+2j,l}$ mutually commute, which indicates that they may gap out the bulk in a translation invariant manner \cite{Kane2002Jan}. 

We systematically searched for the most relevant instability on a finite system of $N=300$ wires to generate the phase diagram  Fig. \ref{fig:crystals}b. At each set of microscopic parameters we tested, the Chern number agreed among the most relevant instabilities. The phase diagram reveals the formation of Hall crystals with various Chern numbers as filling and moir\'e period are varied. The system has a particle-hole $\times$ inversion symmetry which manifests as a symmetry under $\nu\to 1-\nu$, $C\to 1-C$. In Fig. \ref{fig:crystals}c we visualize the mean-field charge density profile at $\Phi_0 = 3/2, \nu=1/3$ by solving the noninteracting problem with both $U_2 = 0$ and $U_2/U_1 = 0.2$, a heuristic depiction of ``before and after" interactions are switched on. 
\par 
When $B/B_0  = (a_0/\ell)^2\gg 1$, modes are well-localized to the wires and $C=0$ or $C=1$ phases dominate. These both arise from backscattering between nearest neighbor pairs as in Fig. \ref{fig:wires}, and correspond to $j_1=j_2=1$ and $I_0$ odd (even) for the $C=0$ ($C=1$) state. The $C=1$ state has an unpaired chiral mode on a system with left or right termination. 
\par 
As $B$ decreases, the magnetic length becomes larger and the wavefunctions in several adjacent unit cells strongly overlap, leading to longer-range interactions. Interestingly, the Chern number of the Hall crystal concurrently grows in magnitude. For instance, when $\nu<1/2$, we observe a sequence $C=0,1,-1,2,-2$ for decreasing $a_0/\ell$. We have thus demonstrated by a controlled analysis the possibility of \textit{higher} Hall crystals. This sequence extends, though not exactly, down to even smaller fields \cite{supp}. 

A simple explanation for this is that the system tends to choose its Chern number such that 
\begin{equation}\label{eq:cstar}
    Q \ell \approx c_*,
\end{equation}
where $c_*$ is a nonuniversal constant depending on the interactions. $c_*$ is approximately the location of the maximum of the bosonized density-density interactions. Indeed, we find \cite{supp} that we can approximately reproduce the phase diagram Fig. \ref{fig:crystals}b by minimizing $|Q\ell -c_*|$ for $c_* \approx 1.0 -1.3$. Thus a global feature of these Hall crystals is charge ordering with period $\approx 2\pi\ell/c_*$. \par 

The preceding discussion highlights the importance of a maximum in the forward scattering interactions $\mathsf{U}_{IJ}$ at some interwire separation $\Delta x$. For all realistic repulsive interactions, $\mathsf{U}_{IJ}$ is positive, vanishes as $\Delta x\to \infty$ due to locality, and vanishes as $\Delta x \to 0$ for spin-polarized electrons due to Pauli exclusion. Therefore such a maximum exists generally, and phase diagrams similar to Fig. \ref{fig:crystals}b should arise for generic interactions. We note that many prior studies of coupled wire constructions were phenomenological, omitting microscopic interactions. Moreover, previous works did not include the exchange interaction, which is crucial for stabilizing $C\neq 0,1$ phases.
\par
\begin{figure}
    \centering
\includegraphics[width=0.8\linewidth]{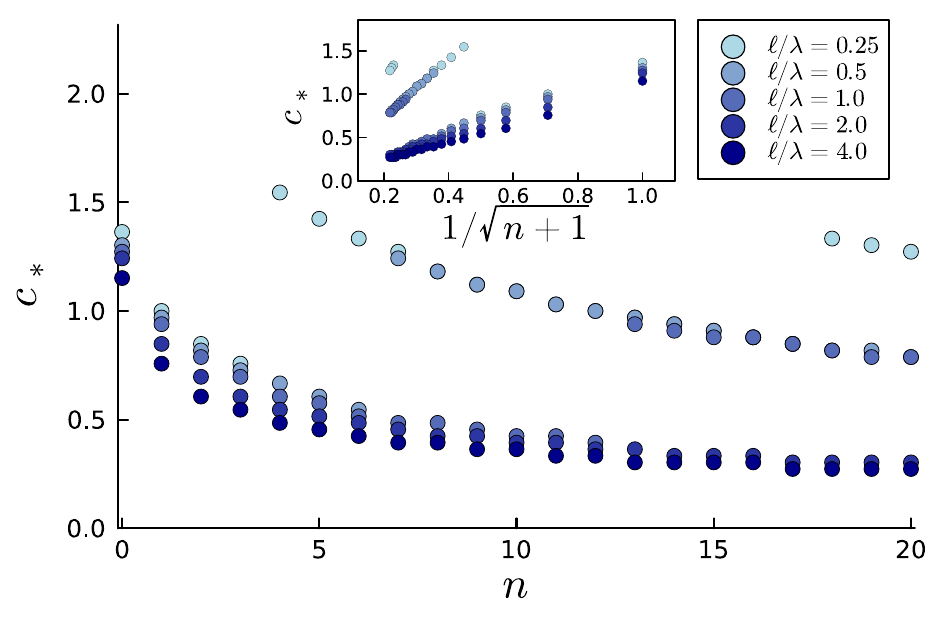}
    \caption{Values of $c_*$ (c.f. Eq. \eqref{eq:cstar}) for Coulomb interactions projected into the $n$'th LL with screening length $\lambda$ ($\ell$ is the magnetic length). Higher hall crystals are modulated with period $\approx 2\pi \ell/c_*$. Inset: plotted over $1/\sqrt{n+1}$. }
    \label{fig:cstar}
\end{figure}

 
The precise value of $c_*$ depends on the form factor $U_{II'JJ'}$, but not on $a_0/\ell$ or $\nu$. Thus we can predict the Hall crystal phase diagram for a wide range of systems simply by specifying $c_*$ and subsequently minimizing $|Q\ell-c_*|$ using $Q=\frac{a_0}{\ell^2} |C - \nu|$. Hence, the single parameter $c_*$ efficiently distills the main content of the phase diagram. In Fig. \ref{fig:cstar}, we provide values for $c_*$ for a range of Landau level indices $n$ and screening lengths $\lambda$ (leaving explicit form factors to \cite{supp}). We observe $c_*\sim 1/\sqrt{n}$ holds to good approximation, indicating the cyclotron radius $r_c \sim \sqrt{n}\ell$ is the length scale of $y$ modulations for higher Landau levels. For long range interactions (large $\lambda$), the interaction has competing local maxima, leading to several ``branches" of $c_*$, the origins of which are the nodes of higher LL wavefunctions. These nodes also lie at the root of the stripe and bubble phases in higher LLs \cite{Fogler1996Jul,Moessner1996Aug}, where the characteristic wavelength is also $\sim r_c$. However, these phases are not adiabatically connected in general due to differing Chern numbers.

\textit{Discussion.---}
In summary, we have proposed an experimentally accessible platform to realize Hall crystals with various Chern numbers, namely a two-dimensional semiconductor or graphene in a strong field and one-dimensional modulation. Hall crystals combine aspects of crystalline order and quantum Hall physics into an intriguing new phase of matter.  While (anomalous) Hall crystals with unit Chern number have been observed in the conventional Hall setting (and possibly in moir\'e systems), Hall crystals with higher or ``sign-flipped" Chern number, driven by the interaction-induced crystallization itself, have not been previously theorized, nor observed. These may host new phenomena beyond the usual purview of quantum Hall physics. 
\par 
In particular, a patterned dielectric substrate with a 1d superlattice \cite{Forsythe2018Jul,Li2021May,Li2021Dec} may help to unlock these phases. The first signatures of these states occur below $a_0/\ell \sim 3$, which corresponds to $B\sim$ a few tesla assuming a realistic $a_0 \sim$ 35 nm \cite{Forsythe2018Jul}.
Of course, the relevant experimental platforms may deviate in a few ways from the pristine model considered in this work. First, the superlattice may only have an approximate periodicity $a_0$ due to strain variations or patterning imprecision. In this case, many instabilities considered herein would fail to conserve the dipole symmetry at strong fields, and would hence be rendered irrelevant by rapidly oscillating phases at large distances, increasing likelihood of a stable smectic metal. Indeed, if $V(x)$ is the sum of incommensurate harmonics, there would be \textit{no} relevant instabilities. It is interesting to ask how much strain variation the CDW can survive before giving way to the smectic metal, given that it is an extended phase and should survive infinitesimal perturbations. Second, defects, disorder, or an underlying $y$ modulation may spoil the translation symmetry. In this case fermion bilinears $\psi_I^\dagger \psi_J$ are the most relevant instabilities and the system likely flows to a 2D Fermi liquid. The temperature scale for this melting, however, is exponentially small in the perturbation strength. Third, higher harmonics $\cos(mq_0 x)$ may be present in $V(x)$. However, their contributions are strongly suppressed in a $B$ field by $e^{-m^2q_0^2\ell^2/4}$, and hence they typically do not alter the physics described here. Strict adherence to the profile of $V(x)= V_0\cos(q_0x)$ is certainly not necessary.\par 
Moreover, a physical magnetic field is not a requirement. Strain engineering in graphene can produce pseudomagnetic fields of order $100$ T \cite{Levy2010Jul}, whose alternating pattern naturally gives rise to an array of one-dimensional ``snake'' states \cite{Tang2014Dec,Venderbos2016May}. In our context, these could serve as chiral modes whose interactions may drive (anomalous) Hall crystal formation. Beyond strained graphene, an interesting future direction is the extension of the current work to more general Chern bands. 

Designing and observing (higher) Hall crystals are compelling goals, and our work suggests a direct route towards this through 2D material engineering. Furthermore, their phenomenology presents a fascinating direction for future studies, particularly at strong interaction, where competition with FQH phases \cite{Kane2002Jan,Teo2014Feb,Shavit2024May} becomes important.
\par 
\textit{Acknowledgments.---}  We thank Pak Kau Lim, Valentin Crepel, and Bert Halperin for helpful discussions. NP acknowledges the KITP where this work was initiated. This research was supported by the Air Force Office of Scientific Research (AFOSR) under award FA9550-22-1-0432. This research was supported in part by the Heising-Simons Foundation, the Simons Foundation, and grants no. NSF PHY-2309135 to the Kavli Institute for Theoretical Physics (KITP). GS acknowledges support from the Walter Burke Institute for Theoretical Physics at Caltech, and from the Yad Hanadiv Foundation through the Rothschild fellowship. LF was supported in part by a Simons Investigator Award from the Simons Foundation.

\bibliography{bib}

\end{document}


\title{Supplemental Material: Designing (higher) Hall crystals}

\author{Nisarga Paul}
\affiliation{Department of Physics, Massachusetts Institute of Technology, Cambridge, MA,
USA}
\author{Gal Shavit}
\affiliation{Department of Physics and Institute for Quantum Information and Matter,
California Institute of Technology, Pasadena, CA, USA}
\affiliation{Walter Burke Institute of Theoretical Physics, California Institute of Technology, Pasadena, CA, USA}
\author{Liang Fu} 
\affiliation{Department of Physics, Massachusetts Institute of Technology, Cambridge, MA,
USA}

\maketitle
\onecolumngrid

\tableofcontents

\section{Details of bosonization procedure}\label{app:bos}
The microscopic Hamiltonian we study, after projecting interactions into the lowest Landau level and accounting for dispersion, is 
\begin{subequations}
    \begin{align}
        H &= H_0 + H_1 \\
        H_0 &= \sum_{k} E_0(k) c_{k}^\dagger c_{k} \\
        H_1 &= \sum_{k_1,k_2,k_3,k_4} U_{k_{1},\ldots, k_{4}} c_{k_{1}}^\dagger c_{k_{2}}^\dagger c_{k_{3}} c_{k_{4}} 
    \end{align}
\end{subequations}
where $c_{k}^\dagger$ (here $k=k_y$) creates a spinless fermion in a Landau gauge orbital
\begin{equation}
    \varphi_{0,k}(\vec r) = \frac{1}{\sqrt{L_y\ell}} e^{-ik y}e^{-(x-k\ell^2)^2/2\ell^2}
\end{equation}
with canonical anticommutation relations $\{ c_{k},c_{k'}\} = 0,\{ c_{k}^\dagger,c_{k'}\} = 2\pi\delta(k-k')$. The dispersion is $E_{0}(k) =  \frac12 \hbar \omega_c+ \sum_i V_i e^{-q_i^2\ell^2/4} \cos(q_i k \ell^2)$ for a modulation $V(x) = \sum_i V_i \cos(q_ix)$. $A_{\{k_{i}\}}$ captures Coulomb interactions, which we detail later. We work in the regime $|A|\ll |V_i|\ll \omega_c$ so that the low-energy degrees of freedom at generic fillings $\nu$ are chiral edge modes located at some set of positions which we label as
\begin{equation}
    \{\text{wire $x$ positions}\} = \{x_I\} = \{\ell^2 k_{y,I}\}
\end{equation}
for $I\in \mathbb{Z}$. $L_y$ is the system size in the $y$ direction and $\ell = 1/\sqrt{B}$ ($\hbar = e=1$).\par 

Now, let us specialize to the case of a cosine potential $V(x) = V_0 \cos(q_0 x)$ with $V_0>0$. Let $a_0=2\pi/q_0$. Actually, the following applies to any inversion-symmetric potential with a single minimum per unit cell located at $a_0/2$ and is easy to generalize. The wires are located at $y$-momenta
\begin{equation}
    k_{y,I} = \frac{a_0}{\ell^2} \left(\left\lfloor \frac{I-1}{2}\right\rfloor + \frac12 +(-1)^I\frac{\nu}{2}\right)
\end{equation}
or, alternatively, at $x$ positions $x_I = k_{I}\ell^2$. For a system of infinite $x$ extent, $I$ is any integer. For a system with a boundary on the left, we will restrict to $I > 0$, and for a system with a boundary on the right, we will restrict to $I\leq 0$. For a system of finite extent, we restrict to $I = \{1,\ldots, 2N\}$ where $N \in \mathbb{Z}$. These are all conventions that do not lose generality. The velocities are
\begin{equation}
    v_I = \pdv{E(k)}{k}\left.\right|_{k = k_{I}} \equiv (-1)^I v
\end{equation}
where $v>0$. We can imagine a system of finite extent as a potential $V(x)$ going to $+\infty$ at both edges; then it is clear that there are an even number of wires and the chirality at each edge is unambiguous. For a modulation $V(x) = V_0 \cos(2\pi x/a_0)$, the velocity is 
\begin{equation}
    v = \frac{2\pi V_0  \ell^2}{a_0} e^{-\pi^2\ell^2/a_0^2} |\sin(\pi \nu)|.
\end{equation}
\begin{figure}
    \centering
\includegraphics[width=0.8\linewidth]{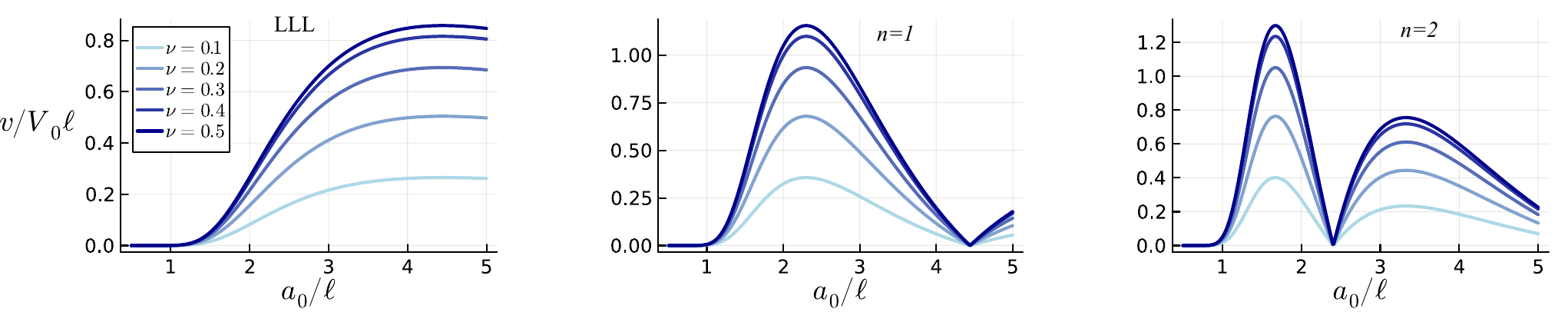}
    \caption{Velocities $v$ in coupled wire model for various $\nu$ and $a_0/\ell$ and for the LLL, $n=1$ LL, and $n=2$ LL, respectively.}
    \label{fig:vels}
\end{figure}
We've plotted this and its analogs in the $n=1$ and $n=2$ Landau levels in Fig. \ref{fig:vels}. The fermionic operator $\psi(y) = \int \frac{dk}{2\pi} c_{k} e^{-ik y}$ satisfies $\{\psi(y),\psi(y')\} = 0,\{\psi(y),\psi(y')^\dagger\} = \delta(y-y')$. Confining ourselves to the vicinities of the wires, we decompose this operator into contributions from each wire:
\begin{equation}
    \psi(y) \simeq \sum_I \int_{k_{y,I}-\Lambda}^{k_{I}+\Lambda} \frac{dk}{2\pi} e^{-i ky}c_{k} \simeq \sum_I \psi_I(y)
\end{equation}
where $\Lambda\ll a_0/\ell^2$ is a momentum cutoff. Linearizing the dispersion around the $k_{I}$ by writing $E_{k} \approx E_F + v_I k$ near the $I$'th wire, we have in the Heisenberg picture $[\partial_t + v_I \partial_y]\psi_I(y)\approx 0$. The approximation becomes exact as $\Lambda \to 0$. The Hamiltonian restricted to these wires is 
\begin{subequations}
    \begin{align}
        H &= H_0 + H_{\text{int}}\\
        H_0 &= -i\sum_I \int dy \, v_I \psi_I^\dagger \partial_y \psi_I\\
        H_{\text{int}} &=  \sum_{I,I',J',J} \int dy \, U_{I,I',J',J} \psi_I^\dagger \psi_{I'}^\dagger \psi_{J'} \psi_J
    \end{align}
\end{subequations}
where $U_{I,I',J',J} = L_y U_{k_{I},k_{I'},k_{J'},k_{J}}$. We introduce chiral bosons $\phi_I$ via the standard dictionary
\begin{equation}\label{eq:psiIdag}
    \psi_I^\dagger = (2\pi\alpha)^{-1/2} e^{-ik_{y,I}y} e^{\pm i\phi_I} \gamma_I
\end{equation}
where $\alpha \sim 1/\Lambda$ is a short-distance cutoff and the $\gamma_I$ are Klein factors satisfying $\gamma_I \gamma_J = -\gamma_J \gamma_I$. The sign in the exponent is $(-)^{I+1}$, i.e. $\psi_I^\dagger \sim e^{-i\phi_I}$ when $v_I > 0$ and $\psi_I^\dagger \sim e^{+i\phi_I}$ when $v_I <0$.
\par 
The bosonized action takes the form described in the main text, 
\begin{subequations}
    \begin{align}
        S &= S_0 +  S_{1} \\
        S_0 &= \frac{1}{4\pi} \int dtdy\, [\mathsf{K}_{IJ} \partial_t \phi_I \partial_y \phi_J - \mathsf{V}_{IJ} \partial_y \phi_I \partial_y \phi_J],
    \end{align}
\end{subequations}
where $\mathsf{K}_{IJ} = (-1)^I\delta_{IJ}$ and $\mathsf{V}_{IJ} = v \delta_{IJ} + (1/\pi)\mathsf{U}_{IJ}$ and $\mathsf{U}_{IJ} = U_{I,J,J,I}-U_{I,J,I,J}$ captures forward-scattering (all terms which become density-density after bosonization). The remaining parts of $H_{\text{int}}$ become cosine operators after bosonization, which are packaged into $S_{1}$.

\subsection{Coulomb interaction}
Here we derive the explicit form of $U_{I,I',J',J}$ used in our calculations. We'll interchangeably write $(I,I',J',J)\sim (1,2,3,4)$ as shorthand. For lowest Landau level wavefunctions, it takes the form
\begin{equation}
    U_{1,2,3,4} = \frac12L_y\int d^2\vec r_1d^2\vec r_2\varphi_{0,1}^*(\vec r_1)\varphi_{0,2}^*(\vec r_2)V_C(\vec r_2-\vec r_1)\varphi_{0,3}(\vec r_2)\varphi_{0,4}(\vec r_1).
\end{equation}
Here, we take a screened Coulomb interaction $V_C(\vec r) = \frac{e^2}{\epsilon r}e^{-r/\lambda}$, where $\epsilon$ is the dielectric constant and $\lambda$ the screening distance. We wish to find 
\begin{equation}\label{eq:A1234}
U_{1,2,3,4}=\frac{1}{2L_y\ell^2} \frac{e^2}{\epsilon } \int d^2\vec r_1d^2\vec r_2 \,e^{ik_{41} y_1} e^{ik_{32} y_2} \frac{e^{-\sqrt{x_{12}^2 + y_{12}^2}/\lambda}}{\sqrt{x_{12}^2+y_{12}^2}} e^{-\frac{(x_1-k_4\ell^2)^2 + (x_2-k_3\ell^2)^2+(x_2-k_2\ell^2)^2+(x_1-k_1\ell^2)^2}{2\ell^2}}.
\end{equation}
\begin{figure}
    \centering
\includegraphics[width=0.8\linewidth]{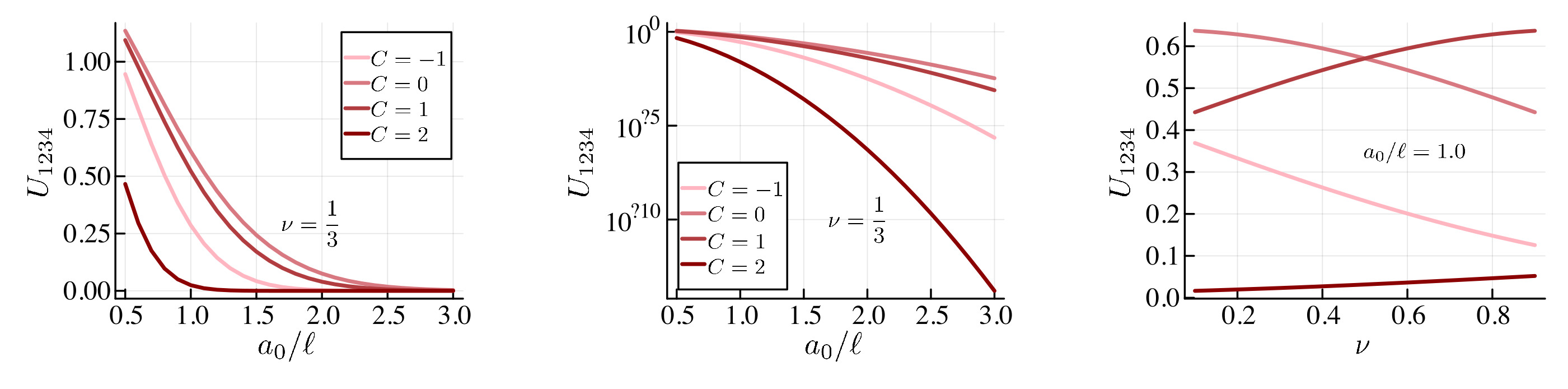}
    \caption{Bare values of $U_{1,2,3,4}$ (Eq. \eqref{eq:A1234}) in arbitrary units for $\ell=\lambda$. (Left, middle) As a function of $a_0/\ell$ at fixed $\nu=1/3$ on a regular and log axis, and (right) as a function of $\nu$ at fixed $a_0/\ell$. }
    \label{fig:barevalues}
\end{figure}
We let $y_{12} = y$ and $y_1+y_2 =Y$. The $y_i$ integration yields
\begin{subequations}
    \begin{align}
        &\int\frac12 dy dY e^{\frac{i}{2}p'y} e^{\frac{i}{2}(k_{41}+k_{32})Y} \frac{e^{-\sqrt{x_{12}^2 + y^2}/\lambda}}{\sqrt{x_{12}^2+y^2}} \\
        &= \delta_k \int dx \,\left[\int \frac{dp}{2\pi} e^{-ip(x'-x)}\right] \int dy' e^{ip'y'} \frac{e^{-2\sqrt{x^2+y'^2}/\lambda}}{\sqrt{x^2+y'^2}}\\
        &= \delta_k \int \frac{dp}{2\pi} e^{-ipx'} \int dr\, 2\pi J_0(\sqrt{p^2+p'^2}r) e^{-2r/\lambda}\\
        &= 2\delta_k K_0\left(\frac{x'}{2}\sqrt{p'^2+4\lambda^{-2}}\right)
    \end{align}
\end{subequations}
where we wrote $\delta_{k}\equiv 2\pi\delta(k_{41}+k_{32})$ (with $\delta_{0} \equiv L_y$) and set $p' = k_{41}-k_{32}$, $x'=x_{12}/2,y'=y/2$ in intermediate steps. Next, let $x_{12} = x$ and $x_1+x_2 = X$, so $x_1 = (X+x)/2$ and $x_2 = (X-x)/2$. We introduce the parameterization 
\begin{equation}
    k_1 = k-q/2,\quad k_2 = k+q/2,\quad k_3 = k-p/2,\quad k_4 = k+ p/2
\end{equation}
which enforces momentum conservation. The integral over $X$ yields 
\begin{equation}
   \int dX e^{-\frac{(x_1-k_4\ell^2)^2 + (x_2-k_3\ell^2)^2+(x_2-k_2\ell^2)^2+(x_1-k_1\ell^2)^2}{2\ell^2}} = \sqrt{2 \pi } \ell e^{ -\frac{\ell^2}{4}
\left(p^2+q^2\right)-\frac{ x^2}{2\ell^2}+\frac{x}{2} (p-q)}.
\end{equation}
Altogether, 
\begin{equation}
U_{1,2,3,4} = \frac{\sqrt{2\pi}e^2}{\epsilon }  e^{-\frac18 \ell^2(p+q)^2}\int ds\, K_0\left(\frac{\ell}{2}\sqrt{(p+q)^2 + 4\lambda^{-2}}|s+\ell(p-q)/2|\right) e^{-s^2/2}.
\end{equation}
We plot representative examples of $U_{1,2,3,4}$ in Fig. \ref{fig:barevalues} for a few different Chern numbers $C$. These give a guide to the eye for the relative weights of the bare couplings of various CDW terms in the microscopic Hamiltonian. We associate values of $q\ell$ and $p\ell$ to Chern numbers through Fig. 2b of the main text; for instance, $C=2$ refers to $p\ell = (a_0/\ell)(4-\nu)$ and $q\ell = (a_0/\ell)\nu$. It is clear that $C=0,1$ are generally the most dominant in bare values, though the other CDWs increase in bare value for smaller $a_0$. \par 
The density-density interactions satisfy 
\begin{subequations}
\begin{align}\label{eq:uijdd}
    \mathsf{U}_{IJ} &= \frac{\sqrt{2\pi}e^2}{\epsilon} \tilde U(q\ell,\ell/\lambda),\\
    \tilde U(x,y) &=- \sqrt{\frac{\pi }{2}} e^{-(x^2-y^2)/4} K_0\left((x^2+y^2)/4\right)+\int ds K_0\left(|s+x|y\right) e^{-s^2/2} 
\end{align}
\end{subequations}
where $q = k_{y,I}-k_{y,J}$ is the momentum separation between wires and $K_0$ is the modified Bessel function of the 2nd kind. We will call the dimensionless ratios $\mathsf{U}_{IJ} / |v_I|$ the ``wire-wire couplings." These would be the $g$ parameters in the literature on Luttinger liquids. \par
\begin{figure}
\includegraphics[width=0.3\linewidth]{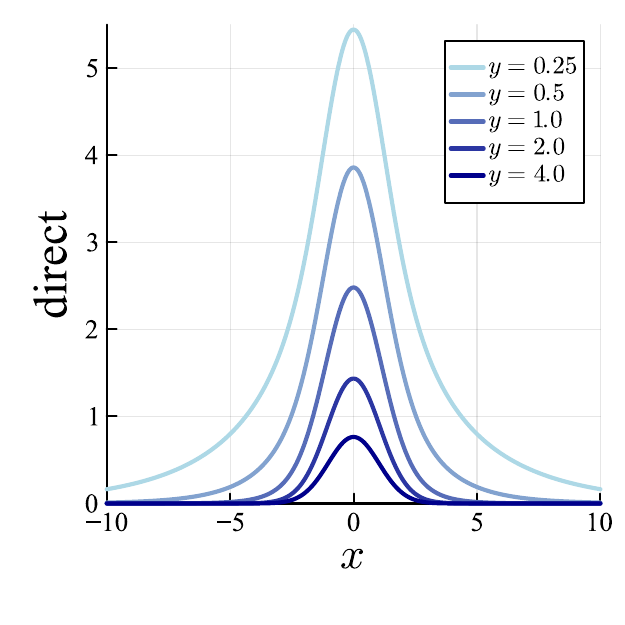}
\includegraphics[width=0.3\linewidth]{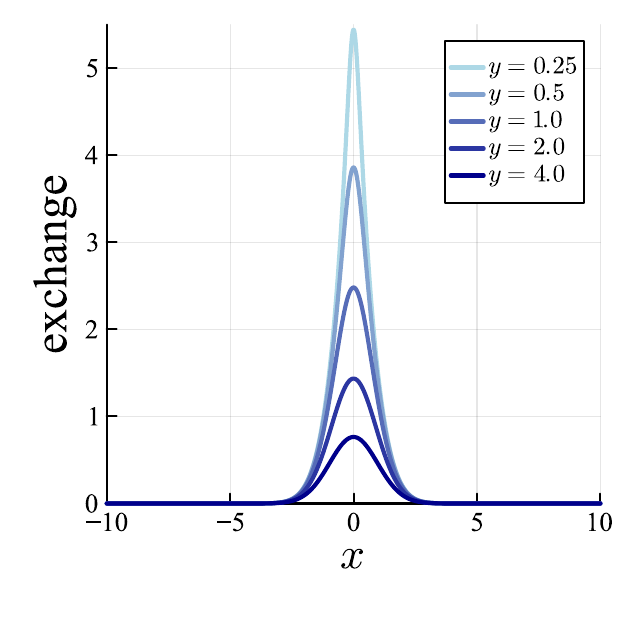}
\includegraphics[width=0.3\linewidth]{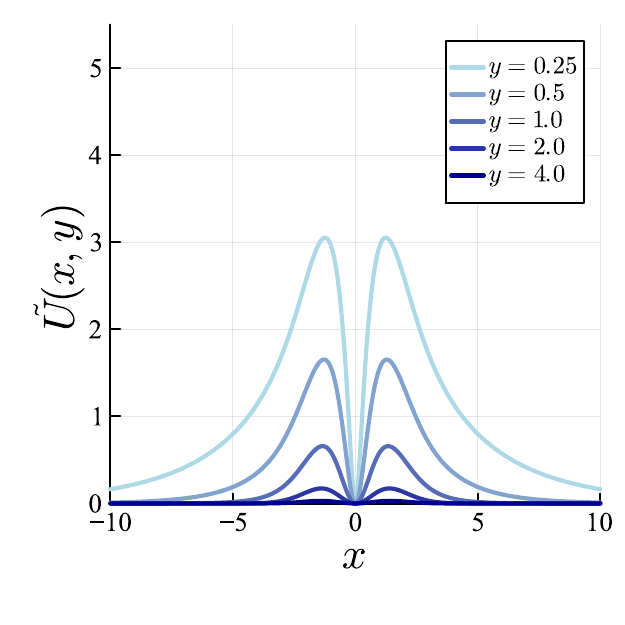}
\caption{(Left to right) The direct, exchange, and total bosonized density-density interaction. The total is given by $\tilde U(x,y)$ = direct - exchange, where $\tilde U(x,y)$ is defined in Eq. \eqref{eq:uijdd}.}
\label{fig:supp1}
\end{figure}

 
There are three length scales involved: $1/q,\lambda,$ and $\ell$. $\tilde U(x,y)$ captures the dependence on two ratios $q\ell, \ell/\lambda$. In Fig. \ref{fig:supp1} we plot several instances of $\tilde U(x,y)$. The integrated strength of the Coulomb interaction diverges for large $\lambda$. Since ours is a perturbative stability analysis, and the important effect of tuning $\lambda$ is in altering the profile of the wire-wire couplings, we normalize the set of couplings $\mathsf{U}_{IJ}/v_{\text{min}}$, where $v_{\text{min}} = \text{min}(|v_I|)$, by setting its maximum value to $g_* = 0.05$ for each $\lambda$. Equivalently, we scale $e^2/\epsilon$ down with increasing $\lambda$ in order that the maximum wire-wire coupling remains fixed. We have checked that adjustments to $g_*\in(0,0.1]$ do not appreciably alter the phase diagrams presented. Beyond $g_*=0.1$, we find some phase transitions to new Hall crystals (in limited portions of the phase diagram, with with even higher Chern numbers possible). However, an analysis of the full RG flow is necessary to confirm their existence.

\subsection{Higher Landau levels}
\begin{figure}
    \centering
\includegraphics[width=0.3\linewidth]{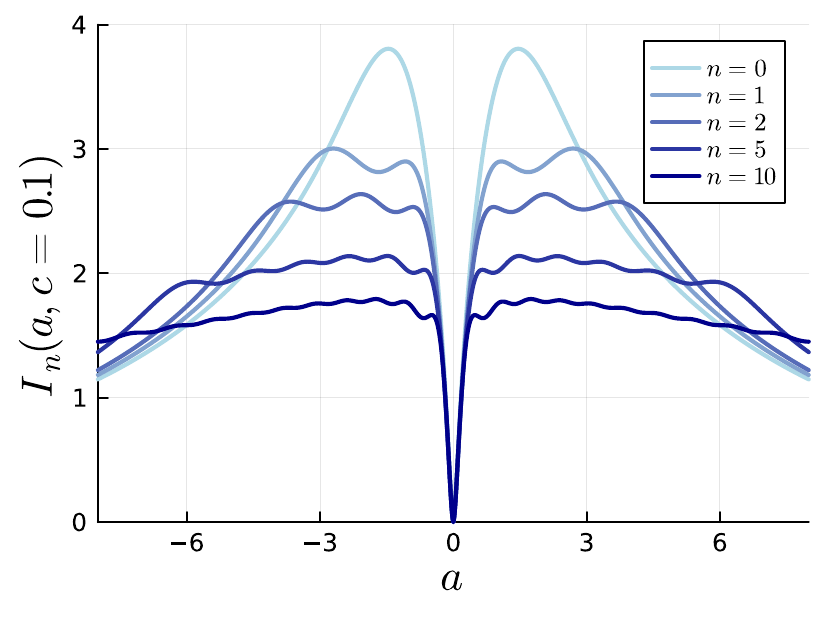}
\includegraphics[width=0.3\linewidth]{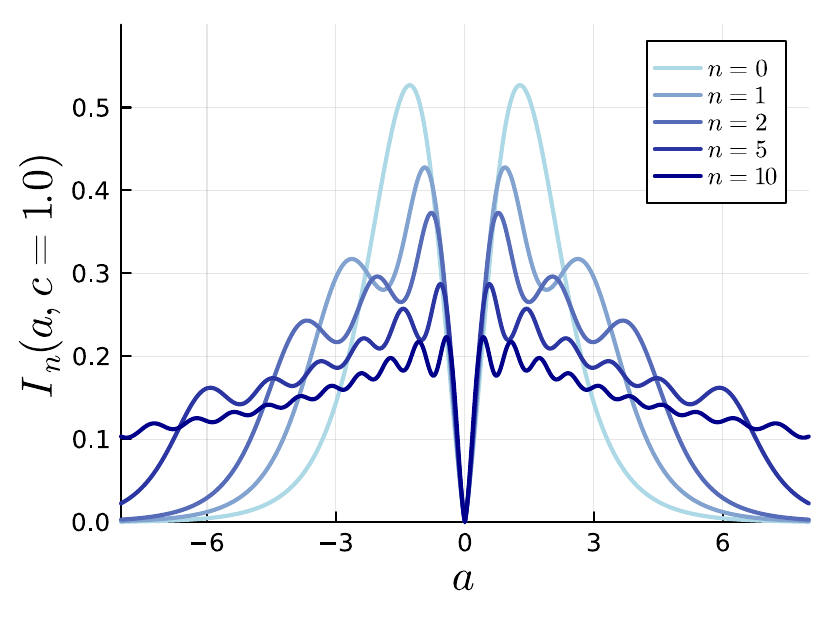}
\includegraphics[width=0.3\linewidth]{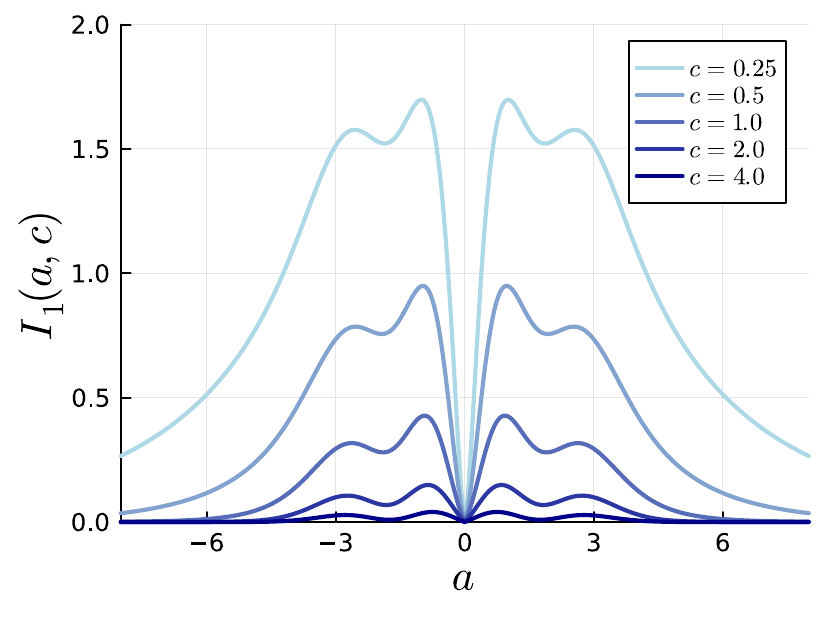}    \caption{Eq. \eqref{eq:Iac} plotted for various $n$ with $c=0.1$ (left) and $c=1$ (middle) and for various $c$ with $n=1$ (right).}
    \label{fig:Iac}
\end{figure}

Going through a similar calculation for higher Landau levels (index $n$) results in 
\begin{equation}
    U_{1,2,3,4} = \frac{e^2}{\epsilon} \tilde I_n(p\ell,q\ell,\ell/\lambda)
\end{equation}
where we've defined a dimensionless integral 
\begin{multline}
\tilde I_n(a,b,c) = C_n^4\int dS\, ds \, H_n((S+s-a)/2) H_n((S-s+a)/2) H_n((S-s-b)/2) H_n((S+s+b)/2) \\
\times K_0\left(\frac{s}{2}\sqrt{(a+b)^2+4c^2}\right)e^{-\frac{(S+s-a)^2 + (S-s+a)^2+(S-s-b)^2+(S+s+b)^2}{8}}
\end{multline}
where $C_n=\frac{1}{\sqrt{\sqrt{\pi}2^nn!}}$. The density-density interaction between two wires corresponds to the $p = -q$ contribution ($k_1 = k_4, k_2 =k_3$) minus the $p=q$ contribution ($k_1 = k_3, k_2 = k_4$). In other words, letting $p = |k_I-k_J|$ be the momentum separation between the two wires, we have $\mathsf{U}_{IJ} = \frac{e^2}{\epsilon} I_n(|k_I-k_J|\ell,\ell/\lambda)$ where
\begin{equation}\label{eq:Iac}
    \tilde I_n(a,c) = \tilde I_n(|a|,-|a|,c)-\tilde I_n(|a|,|a|,c).
\end{equation}
We plot examples in Fig. \ref{fig:Iac}. 

\section{Properties of strong coupling phases}

In this section, we discuss various properties of the strong coupling CDW phases. Such phases are realized by the growth of a relevant operator $\mO_{\vec m} = \cos(m_I  \phi_I)$ (and its translated copies) under RG flow. $\mO_{\vec m}$ must respect the charge and $k_y$ conservation of the microscopic model. This allows us to restrict to four-fermion operators satisfying 
\begin{equation}
    \sum_I m_I (-1)^I = \sum_I m_I (-1)^I k_{y,I} =0.
\end{equation}
The natural translation operation on the operators by one period $a_0$ acts as $T(\mO_{\vec m}) = \mO_{t(\vec m)}$ where $t(\vec m)_I = \vec m_{I-2}$ (with indices modulo $2N$). For charge and $k_y$ conserving operators, $T^j(\mO_{\vec m})$ commute ($t^j(\vec m)^T\mathsf{K} \vec m = 0$)  for all $j$ and can therefore gap out the bulk in a translationally invariant manner. \par 
Note that all symmetry-allowed four-fermion operators are of the form 
\begin{equation}\label{eq:I0j1j2}
\mO_{I_0,J_0,l} = \int dy \, \psi_{J_0+l}^\dagger \psi_{I_0 - l}^\dagger \psi_{I_0}\psi_{J_0} +\text{H.c.}.
\end{equation}
The bosonized form of the above is $\mO_{\vec m}$ with 
\begin{equation}\label{eq:mvalues}
    (\mathsf{K}\vec m)_I = -\delta_{I,I_0} - \delta_{I,J_0} + \delta_{I,J_0+l}+\delta_{I,I_0-l}.
\end{equation}
What is the density profile and Chern number of the CDW corresponding to $\mO_{\vec m}$? The density operator is 
\begin{equation}
   \rho(y)= \psi^\dagger(y) \psi(y) \simeq \sum_{I,J} \psi_I^\dagger(y) \psi_J(y)
\end{equation}
or in bosonized form, 
\begin{subequations}\label{eq:rho}
\begin{align}
    \rho(y) &= -\frac{1}{\pi}\sum_I\partial_I\phi_I(y) + \frac{1}{2\pi\alpha} \sum_{I\neq J}\kappa_I\kappa_Je^{i(k_{y,I}-k_{y,J})y} e^{-i((-)^{I}\phi_I(y)-(-)^{J}\phi_J(y))}\\
    &= -\frac{1}{\pi}\sum_I\partial_I\phi_I(y) + \frac{1}{\pi\alpha} \sum_{I< J}\sin((k_{y,I}-k_{y,J})y)\cos(\vec n^{(I,J)}\cdot \vec \phi)-\cos((k_{y,I}-k_{y,J})y)\sin(\vec n^{(I,J)}\cdot \vec \phi)
\end{align}
\end{subequations}
where in the last step we chose a simultaneous $-i$ eigenstate of the Klein factors and defined a boson
\begin{equation}\label{eq:nIJ}
     \vec n^{(I,J)} \cdot \vec\phi \equiv (-)^I\phi_I-(-)^J\phi_J.
\end{equation}
If $\vec n^{(I,J)} \cdot \vec\phi$ acquires long-range order (LRO), the density acquires a modulation with wavevector $k_{y,I}-k_{y,J}$. This can be determined by localizing the $t^j(\vec m) \cdot \vec \phi$ and replacing $g\mO_{\vec m}$ with $-g \mathsf{L}[\vec m]_{IJ} \phi_I \phi_J/2$, where 
\begin{equation}
    \mathsf{L}[\vec m]_{IJ} = \sum_j t^j(\vec m)_I t^j(\vec m)_J,
\end{equation}
to obtain a quadratic action which can be solved exactly. Using this quadratic action, the boson with
\begin{equation}\label{eq:nIJ2}
    \vec n^{(I,J)} = \begin{cases}
        \vec n^{(I_0, I_0+j_1+j_2)} & j_1 \text{ even}\\
        \vec n^{(I_0,I_0+j_1)} & j_1 \text{ odd}
    \end{cases}
\end{equation}
and its translated copies acquire LRO, and no others. Put another way, the operator $\mO_{\vec m}$ at mean field decouples in the $\psi_{I_*}^\dagger \psi_{J_*}$ channel where $I$ and $J$ are a pair of opposite-chirality indices of fermions operators in $\mO_{\vec m}$. (Strictly speaking, for finite $N$, the vertex operator $e^{i\vec n^{(I,J)}\cdot \vec \phi}$ acquires a power-law decay $1/|y|^{1/N}$, consistent with the Coleman-Mermin-Wagner theorem and signaling a transition to true LRO in the thermodynamic limit). This implies that at long distances the CDW associated to $\mO_{\vec m}$ has a wavevector
\begin{equation}\label{eq:km1}
    Q= \begin{cases}
        |k_{J_0+l}-k_{I_0}| & l \text{ even}\\
        |k_{I_0}-k_{I_0-l}| & l \text{ odd}
    \end{cases}.
\end{equation}
along $y$. Recalling the form of $k_{I}$, Eq. \eqref{eq:km1} can be evaluated and takes the form shown in Table I (main text). In all these cases, we can adopt the suggestive notation 
\begin{equation}\label{eq:km}
    Q= \frac{a_0}{\ell^2} \eta(\nu-C) 
\end{equation}
where $C \in \mathbb{Z}$ and $\eta=\pm 1$. The 2D unit cell has area
\begin{equation}\label{eq:Auc}
    A_0 = \frac{2\pi}{Q} a_0 = \frac{2\pi \ell^2}{\eta(\nu-C)}. 
\end{equation}
The so-called Diophantine relation relates density, field, and $A_0$ in crystalline Hall systems. We have
\begin{equation}
    \bar\rho = C \rho_{\Phi} + \eta_s A_0^{-1},
\end{equation}
where $\bar \rho = \nu/2\pi \ell^2$ is the 2D electron density and $\rho_{\Phi} = 1/2\pi\ell^2$ is the density of a filled Landau level, and $C,\eta_s$ are integers (or fractions) labelling topological invariants of the state. This is an interacting generalization of the Diophantine equation which determines the Chern numbers in the gaps of the Hofstadter spectrum. Crucially, there need not be a direct proportionality between the number of magnetic flux quanta and number of electrons per unit cell, and neither of these quantities show any tendency to lock to rational numbers. This is precisely what allows Hall crystals to form over \textit{continuous ranges} of field and density. Using Eq. \eqref{eq:Auc}, we find
\begin{equation}
    C = C_{\vec m},\quad \eta_s = \eta_{\vec m}
\end{equation}
in our case. If the CDW is pinned by an infinitesimal external potential, so the chemical potential lies inside a gap and $A_{\text{uc}}$ is fixed, the Streda formula implies
\begin{equation}
    \sigma_{xy} = \frac{e^2}{h} \pdv{\bar\rho}{\rho_\Phi} = \frac{e^2}{h}C,
\end{equation}
which identifies $C_{\vec m}$ with the Chern number.\par

Another way to determine the Chern number is by counting gapless modes on an edge. Let us define a mode $\vec \rho$, the residual charge, by 
\begin{equation}
    \rho_I = \begin{cases} (-1)^I & t^j(\vec m)_I =0 \,\, \forall j\\
    0 & \text{else}
    \end{cases},
\end{equation}
which commutes with all $t^j(\vec m)$. This consists of all modes ``untouched" by the CDW. While $\vec \rho 
= 0$ in an infinite system, $\vec\rho$ may be nonvanishing on a system with a left or right termination, in which case the number of gapless edge modes is $N_{\text{edge}} = \frac12 \vec \rho^T \mathsf{K} \vec \rho$ and, provided the bulk Goldstone mode is (infinitesimally) gapped, the edge conductance is $G = \frac{e^2}{h} N_{\text{edge}}$. One finds precisely
\begin{equation}
    N_{\text{edge}} = C
\end{equation}
as one would expect from the bulk / boundary correspondence. Moreover, these edge modes are topologically protected: edge modes of opposite chiralities live on opposite edges (so backscattering is suppressed by large separation) while backscattering processes between edge modes and bulk modes are irrelevant.

\section{Additional plots}
\begin{figure}
    \centering
\includegraphics[width=0.3\linewidth]{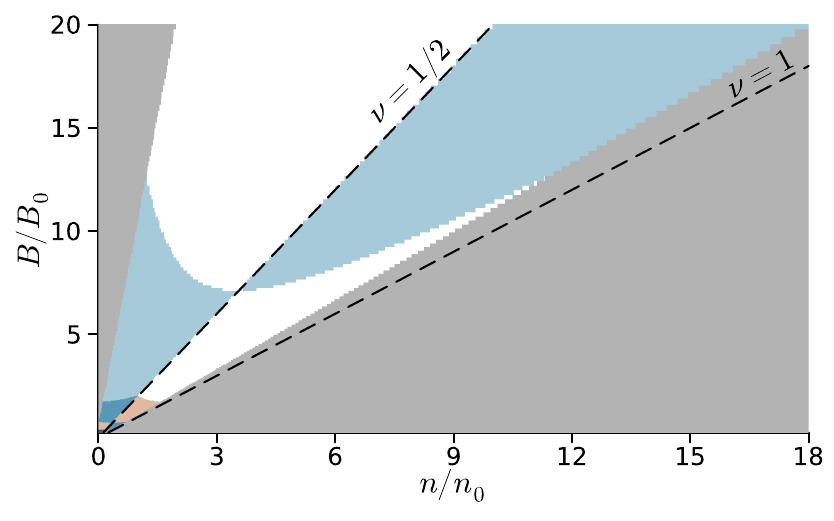}
    \caption{Phase diagram from main text as a function of density in $y$ axis and field in the $x$ axis, where $n_0 = B_0 = 1/2\pi a_0^2$. Chern numbers follow same color coding as before. Gray are regions outside of calculation. }
    \label{fig:density}
\end{figure}
\begin{figure}
    \centering
\includegraphics[width=\linewidth]{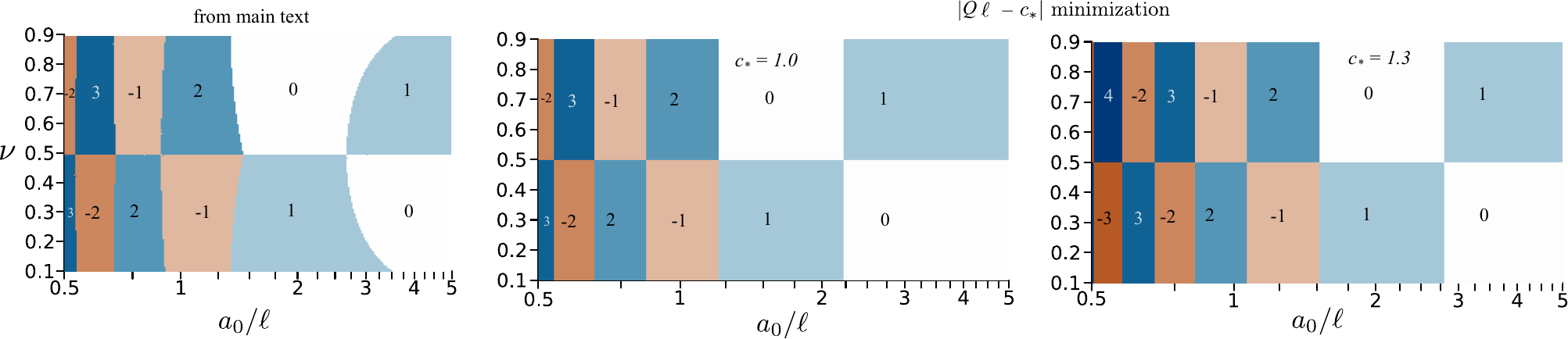}
    \caption{Left: phase diagram from main text. Middle,right: result of minimizing $|Q\ell- c_*|$ for $C$, where $Q$ is given by Eq. \eqref{eq:km}, with $c_* = 1.0$ and $c_*=1.3$.}
    \label{fig:cstarcomparison}
\end{figure}

\textit{Density versus field.} We plot the phase diagram in the lowest Landau level as a function of density in the $y$-axis rather than filling fraction in Fig. \ref{fig:density}. \textit{Chern numbers from minimization protocol. } In the main text, we have noted that the system chooses its Chern number by minimizing $|Q\ell - c_*|$ for some $O(1)$ interaction-dependent (but $a_0,\nu,\ell$-independent) constant $c_*$. We give further support for this claim here. In Fig. \ref{fig:cstarcomparison} we show the result of minimizing $|Q\ell - c_*|$ for $C$ across a range of parameters $a_0/\ell$ and $\nu$. A broad similarity can be observed in this case and in all others that we have checked. Therefore, the essence of the phase diagram can be captured by a single dimensionless parameter $c_*$, which indicates the characteristic length scale (namely $2\pi \ell /c_*$) by which the system would like to spontaneously order. We note, however, that this minimization protocol does not capture all details of the phase diagram, such as curved phase boundaries.

\bibliographystyle{apsrev4-1}
\bibliography{bib}